\documentclass{kluwer}
\usepackage{graphicx}
\begin{document}
\begin{article}
\begin{opening}

\title{Jet-Disc coupling in the accreting black hole XTEJ1118+480}            
\author{Julien \surname{Malzac}\email{malzac@cesr.fr}\thanks{Partially supported by PPARC}}
\institute{Centre d'Etude Spatiale des Rayonnements, (CNRS/UPS/OMP), Toulouse, France} 
\author{Andrea \surname{Merloni}}
\institute{Max-Planck-Institut f\"ur Astrophysik, 
 Garching, Germany}
\author{Andrew C. \surname{Fabian}}
\institute{Institute of Astronomy, 
 Cambridge, 
United Kingdom}                              
\runningtitle{Jet-Disc coupling in XTEJ1118+480}
\runningauthor{J. Malzac et al.}
\begin{ao}
Centre d'Etude Spatiale des Rayonnements,\\
9, Avenue du Colonel Roche,\\
BP4346 , 31028 Toulouse Cedex 4\\
France
\end{ao} 
\begin{abstract} 
We interpret the rapid correlated  UV/optical/ X-ray
variability of XTE~J1118+480 as a signature of the coupling between the
X-ray corona and a jet emitting synchrotron radiation in the optical
band.We propose a scenario in which the jet and the 
X-ray corona are fed by the same energy reservoir where
 large amounts of accretion power are stored
 before being channelled  into either the jet or 
the high energy radiation. This time dependent model reproduces
 the main features of the rapid multi-wavelength variability of
XTE~J1118+480.  
A strong requirement of the model is that the total 
jet power should be at least a few times
larger than the observed X-ray luminosity, implying a radiative
efficiency for the jet $\epsilon_{\rm j} \apprge  3 \times 10^{-3}$.
This would be consistent with the overall low radiative efficiency of the
source. We present independent arguments showing that the jet
probably dominates the energetic output of all accreting black
holes in the low-hard state.
\end{abstract}
\end{opening}
\section{INTRODUCTION}
The  X-ray nova XTE J1118+480, was discovered by the Rossi X-Ray Timing Explorer
({\it RXTE}) All-Sky Monitor ({\it ASM}) on 2000 March 29 
(Remillard et al. 2000) .
 The optical spectrophotometry proved a low mass X-ray binary system
 containing a black hole of at least 6  solar masses
(McClintock al. 2001a, Wagner et al. 2001).
 The interstellar extinction towards
  the source is exceptionally low (Garcia et al. 2000).
 This fact allowed an unprecedented wavelength coverage
 (Mauche et al. 2000; Hynes et al. 2000; 
McClintock et al. 2001b;
 Hynes et al. 2003; Chaty et al. 2003 and references therein).
In  the radio to optical bands,  
a strong non-thermal component was associated with synchrotron
emission from a powerful jet or outflow (Fender et
al. 2001). In the
optical to EUV bands the spectral energy distribution 
 is dominated by a thermal component from the accretion
disc. The X-ray emission consists of a typical powerlaw spectrum with photon
index $\Gamma \sim 1.8$. Such a  
spectrum is generally associated with Comptonisation in the hot inner  part
of the disc or corona. 
During the whole outburst duration, the X-ray properties of 
the source, as well as the presence of strong radio emission,
were typical of black hole binaries in the hard state.

\section{Optical X-ray correlations}

Interestingly, fast optical and UV photometry allowed by the weak extinction, revealed a rapid optical/UV
flickering presenting complex correlations with the X-ray variability (Kanbach et al. 2001; Hynes et
al. 2003, hereafter K01 and H03 respectively). 
This correlated variability cannot be caused by reprocessing 
of the X-rays in the external parts of the disc.
Indeed, the optical flickering occurs on average on shorter
time-scales than the X-ray one (K01), and reprocessing models fail to 
fit  the complicated shape of the X-ray/optical cross correlation 
function (H03).
Spectrally, the jet emission  seems to
 extend at least up to the optical band (McClintock et al. 2001b; Chaty
et al. 2003, hereafter C03), 
although the external parts of the disc may provide an
important  contribution to the observed flux at such
wavelengths.
The jet  activity is thus the most likely explanation for the rapid
observed optical flickering. For this reason, 
the properties of the optical/X-ray correlation 
in XTE J1118+480 might be of primary importance for the understanding 
of the jet-corona coupling and the ejection process.
 
The simultaneous optical/X-ray observations are described at length in
a number of papers (K01; Spruit \& Kanbach 2001; H03; Malzac et
al. 2003, hereafter M03). As discussed in these works, the observations are very challenging
 for any accretion model. The most puzzling pieces of evidence 
are the following:   
(a) the optical/X-ray Cross-Correlation Function (CCF) shows the optical band lagging the X-ray by $~$0.5
s, but with a dip 2-5 seconds in advance of the X-rays (K01); 
(b) the correlation between X-ray and optical light curves 
appears to have timescale-invariant properties: 
the X-ray/optical CCF maintains a similar, but rescaled, shape on
timescales ranging at least from 0.1 s to few tens of sec (M03);
(c) the correlation does not appear to be  triggered by
 a single type of event (dip or flare) in the light curves; instead, as was
 shown by M03, optical and X-ray  fluctuations of very different shapes, amplitudes
 and timescales are correlated in a similar way, such that 
the optical light curve is related to the time derivative of the X-ray
one.
Indeed, in the range of timescales where the coherence is maximum,
the optical/X-ray phase lag are close to $\pi/2$, indicating that the
two lightcurves are related trough a differential relation.
Namely, if the optical variability is representative 
of fluctuations in the jet power output  $P_{\rm j}$, 
the data suggest that the jet power scales roughly like $P_{\rm j} \propto
-\frac{dP_{\rm x}}{dt}$, where $P_{\rm x}$ is the X-ray power.

\section{The energy reservoir model}

Malzac, Merloni \& Fabian (2004, hereafter MMF04) have shown that the
complex X-ray/optical correlations could be understood in terms of an
energy reservoir model.  In this picture, it is assumed that large
amounts of accretion power are stored in the accretion flow before
being channeled either into the jet (responsible for the variable
optical emission) or into particle acceleration/ heating in the
Comptonizing region responsible for the X-rays.  MMF04 have developed
a time dependent model which is complicated in operation and
behaviour. However, its essence can be understood using a simple
analogue: Consider a tall water tank with an input pipe and two output
pipes, one of which is much smaller than the other. The larger output
pipe has a tap on it. The flow in the input pipe represents the power
injected in the reservoir $P_{\rm i}$, that in the small output pipe
the X-ray power $P_{\rm x}$ and in the large output pipe the jet power
$P_{\rm j}$.  If the system is left alone the water level rises until
the pressure causes $P_{\rm i}=P_{\rm j}+P_{\rm x}$.  Now consider
what happens when the tap is opened more, causing $P_{\rm j}$ to
rise. The water level and pressure (proportional to $E$) drop causing
$P_{\rm x}$ to reduce. If the tap is then partly closed, the water
level rises, $P_{\rm j}$ decreases and $P_{\rm x}$ increases. The rate
$P_{\rm x}$ depends upon the past history, or integral of $P_{\rm
j}$. Identifying the optical flux as a marker of $P_{\rm j}$ and the
X-ray flux as a marker of $P_{\rm x}$ we obtain the basic behaviour
seen in XTE\,J1118+480.  In the real situation, we envisage that the
variations in the tap are stochastically controlled by a shot noise
process. There are also stochastically-controlled taps on the input
and other output pipes as well. The overall behaviour is therefore
complex. The model shows however that the observed complex behaviour
of XTE\,J1118+480 can be explained by a relatively simple basic model
involving several energy flows and an energy reservoir.  This simple
model is largely independent of the physical nature of the energy
reservoir. In a real accretion flow, the reservoir could take the form
of either electromagnetic energy stored in the X-ray emitting region,
or thermal (hot protons) or turbulent motions. The material in the
disc could also constitute a reservoir of gravitational or rotational
energy behaving as described above.
\label{sec:model}
In a stationary flow, the extracted power $P_{\rm j}+P_{\rm x}$ would be perfectly
balanced by the power injected, which is, in
the most general case, given by 
the difference between the accretion power and the power
advected into the hole and/or stored in convective motions: $P_i
\simeq \dot M c^2 - P_{\rm adv,conv}$. However, observations of
strong variability on short time scale clearly indicate that the heating and
cooling of the X-ray (and optical) emitting plasma are highly
transient phenomena, and the corona is unlikely to be in complete 
energy balance on short timescales.
We therefore introduce a time-dependent equation governing the
evolution of its total energy $E$:
\begin{equation}
\dot E= P_{\rm i} - P_{\rm j} - P_{\rm x},
\label{eq:enbal}
\end{equation}
and we assume that all the three terms on the right hand side are time dependent.
The optical variability is produced mainly from 
synchrotron emission in the inner part of the jet at distances of a few thousands
 gravitational radii from the hole. We assume that  at any time the optical flux $O_{pt}$ (resp. X-ray flux)
  scales like the jet power $P_{\rm j}$ ( plasma heating power $P_{\rm x}$).
We introduce the instantaneous dissipation rates $K_{\rm j}$ and $K_{\rm x}$~:
\begin{equation}
 P_{\rm j}(t)= K_{\rm j}(t)E(t),  
 \qquad P_{\rm x}(t)= K_{\rm x}(t)E(t), 
\label{def:kx}
\end{equation}
For a specific set of parameters we generate random independent
fluctuations (time series) for $K_{\rm x}$, $K_{\rm j}$ and $P_{\rm
i}$, solve the time evolution of the energy reservoir $E$ and then use
the solution to derive the the resulting optical and X-ray light
curves (see MMF04 for details).

\begin{figure*}
\centerline{\scalebox{0.7}{\includegraphics{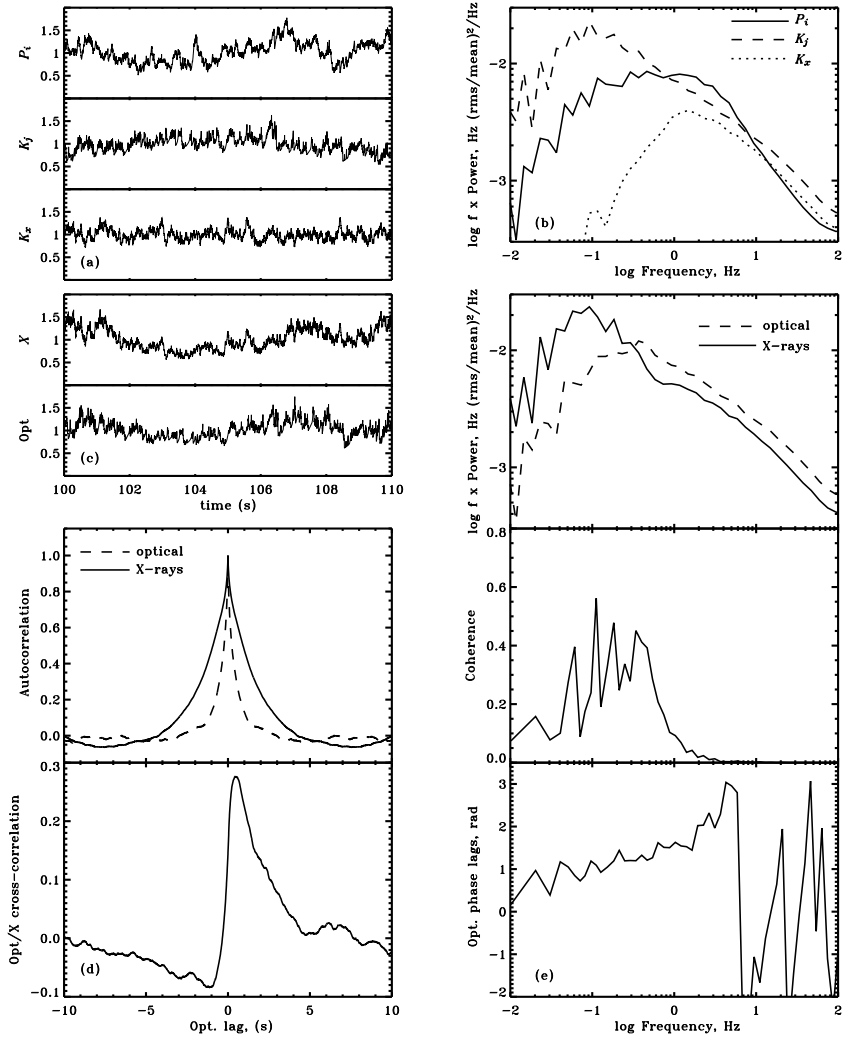}}}

\caption{ Sample input time series (panel a) and power
spectra (panel b) of $P_{\rm i}$,
$K_{\rm j}$, $K_{\rm x}$, resulting X-ray and optical fluxes light curves
(panel c),  X-ray/optical autocorrelation and cross-correlation
functions (panel d), power spectra, coherence and phase-lags (panel e).} 
\label{fig:simu184}
\end{figure*}

Combining equations (\ref{eq:enbal}) and (\ref{def:kx}) we obtain the
following relation for the total instantaneous jet power:
\begin{equation}
P_{\rm j}=P_{\rm i} - (1 + \frac{\dot K_{\rm x}}{K_{\rm x}^2})P_{\rm x} -
\dot P_{\rm x}/K_{\rm x}.
\label{eq:jdom}
\end{equation}
We can see from this equation that the differential
scaling $P_{\rm j} \propto - \dot P_{\rm x}$, observed in
XTEJ1118+480, will be rigorously reproduced  
provided that:
(1 ) $K_{\rm x}$ is a constant;
(2 ) $P_{\rm i} - P_{\rm x}$ is a constant.
It is physically unlikely that those conditions will be  exactly
verified. In particular, $P_{\rm x}$ is observed to have a large RMS
amplitude of variability of about 30 percent. 
However, the observed differential relation holds only roughly
and only for fluctuations within a relatively narrow range of
time-scales $1-10 s$. Therefore, the above conditions need only to be
fulfilled approximatively and for low frequency fluctuations ($> 1$s). 
In practice, the following requirements will be enough to make sure
that  the low frequency fluctuations of the right hand side  
of equation $\ref{eq:jdom}$ are dominated by $\dot P_{\rm x}$:
\begin{itemize} 
\item $P_{\rm x} \ll P_{\rm i}$, implying that the jet power, on average, dominates over the X-ray luminosity;
\item the amplitude of variability of $K_{\rm x}$ and $P_{\rm i}$ in the
$1$-$10$ s range is low compared to that of $P_{\rm j}$. In other words the
$1$-$10$ s fluctuations of the system are mainly driven by the jet activity, 
implying that the mechanisms for dissipation in the jet and the corona occur
 on quite different time-scales.  
\end{itemize}
Figure \ref{fig:simu184} shows the results of a simulation matching
 the main timing properties of XTE J1118+480.
 In this simulation  jet power was set to be 10 times larger than the X-ray power. 
 The model produces an X-ray power spectrum with a plateau
up to $\sim 0.1$ Hz and a power-law 
component with slope $\sim 1.4$ above that frequency, with 
most of the X-ray variability
occurring around 0.1 Hz. The optical PDS power-law has a flatter slope ($\sim 1$)
up to 1 Hz and then softens to a slope similar to
that of the X-ray PDS.
The resulting optical ACF is  significantly narrower than the
X-ray one. The full-width-at-half-maximum (FWHM) of the two ACFs
 differs by a factor $>2$. 
The overall coherence is low ($<0.4$): 
reaching a maximum in the 0.1--1 Hz range and decreasing rapidly
 both at lower and higher frequency. 
The  phase-lags 
are close to $\pi/2$ in the  0.1--1 Hz range and increase from 0 at
low frequencies up to $\pi$ at around 6 Hz. At higher frequencies the
phase lags spectrum is characterized by large oscillations.
 Finally the resulting CCF rises very quickly at positive optical lags,
peaks around $0.5$~s  (this is the post-peak)
and then declines slowly at larger lags.
The two bands appear to be anti-correlated at negative optical
 lags indicating a systematic optical dip 1-2 s before the X-rays
reach their maximum (pre-dip).
All these characteristics are observed in XTE J1118+480.

\section{Jet dominance in XTE J1118+480 and other low-luminosity sources}
\label{sec:flow}
The total mass accretion rate onto the black hole can be estimated
from the observed luminosity of the cold disc component which in the
case of XTE J1118+480 can be estimated from fits to the optical, UV
and EUV spectra (see e.g. Chaty et al. 2003).  The result depends on
several assumptions regarding the geometry and the physical mechanisms
for energy dissipation in the disc and the hot comptonising medium.
However, for reasonable parameters, the estimated accretion rate is
much larger than the observed bolometric luminosity (by at least a
factor of ten).  The important issue, however, would be to determine
whether the missing accretion power escapes the system in the (low
radiative efficiency) jet or in other forms of non-radiative losses,
such as a slow wind, or large scale convective motions, or advection
into the black hole.  The answer to this question resides in the exact
determination of the jet kinetic power.  Unfortunately, there are
major uncertainties in this determination, mainly because the jet
radiative efficiency is not known.  The jet is expected to be a poor
radiator because most of the energy is lost in adiabatic
expansion. Thus, although the radiation from the jet represents a
small fraction of the bolometric luminosity the jet could dominate the
energetics.  For the case of XTE~J1118+480, typical efficiency
$\epsilon_{\rm j}\sim 0.01$ would already imply that the total jet
power dominates over the X-ray luminosity. As discussed above, the
analysis of our time dependent modeling strongly requires $f_{\rm
x}\lesssim 0.1$, corresponding to a jet efficiency $\epsilon_{\rm j}
\lesssim 3\times 10^{-3}$.

There are additional independent arguments in favour of jet dominance 
in low/hard state sources and in XTE~J1118+480 in particular.
Based on the observed radio flux ($L_{\rm R}$) and X-ray correlation observed
in hard states sources (Falcke \& Biermann 1996; Gallo, Fender \& Pooley, 2003),
as well as on standard synchrotron formulae (Heinz \& Sunyaev, 2003), 
Fender, Gallo \& Jonker (2003, hereafter FGP03) have shown that, provided that
advection into the black hole horizon and/or convective motions do not
store a large fraction of the accretion power, there should exist a
critical accretion rate, $\dot m_{\rm cr}$, below which 
an accreting black hole is jet-dominated.
The exact value for the critical accretion rate could be inferred from the
observations, if we knew the total jet power at a
certain X-ray luminosity, and is given by $\dot m_{\rm cr}=2 P_{\rm
  j}^2/L_{\rm x}$, corresponding to a critical X-ray luminosity
$L_{\rm x,cr}=\dot m_{\rm cr}/2$. 
Fender et al. (2001) derived a lower limit for the jet to X-ray power ratio 
in XTE~J1118+480: $P_j/L_{x}=0.2$, and 
FGJ03 used this conservative estimates to determine the value of the
critical rate $\dot m_{\rm cr} \simeq 7 \times 10^{-5}$.
However such a low value of the critical luminosity leads to several
problems.

First,  as shown in FGJ03, during the transition 
from a disc to a jet-dominated state, the dependence of the X-ray luminosity
 on the accretion rate
changes from being $L_{\rm x} \propto \dot m^2$, 
the right scaling for \emph{radiatively
inefficient flows}, to $L_{\rm x} \propto \dot m$, the scaling for
\emph{radiatively efficient flows} (see Fig.~1 of FGJ03).
This would imply that with $L_{\rm x}\sim 10^{-3}$,  XTE~J1118+480
 should be a  \emph{radiatively efficient} system. As discussed above, 
there is however strong observational evidence of the contrary.

Furthermore, black holes in the hard state should show some kind of
spectral transition in the X-ray band at the critical luminosity 
$L_{\rm x,cr} \sim 3 \times 10^{-5}$, due to the drastic changes in
emission mechanisms that are needed to account for the different
scalings of $L_{\rm x}$ with the accretion rate. The observations of
low/hard state sources at such low luminosities are few and hard to
perform, however no indication of any dramatic spectral change in any
hard state source down to quiescent level has ever been reported (Kong
et al., 2002; Hameury et al., 2003)
In fact, the only physical transition that we do actually observe is the
transition between the hard and the soft state that occurs at luminosities
of at least a few percent of Eddington luminosity (Maccarone 2003).
We believe that, if the above mentioned difficulties are to be solved,
then $\dot m_{\rm cr}$ has to
correspond to luminosities that are comparable to, or larger than,
hard-to-soft state transition luminosities. 
For the case of XTE~J1118+480, instead of using the lower limit 
for the jet to X-ray power ($P_j/L_{x}=0.2$), we can adopt 
the much larger value $P_j/L_{x}\sim 10$
required by our variability model. Then we find $\dot m_{\rm cr} \sim 0.2$, involving a transition at
$L_{\rm x,cr}=\dot m_{\rm cr}/2 \sim 0.1$.
 This is in agreement with the idea that the transition from jet
dominated to X-ray dominated states occurs at luminosities similar or
slightly higher than the hard to soft state transition.
Thus, if the arguments of FGJ03 are correct, 
an important consequence of the jet dominance in XTE~J1118+480
is that \emph{all hard state sources are jet-dominated} (in the sense
that the jet power dominates over the X-ray power). 
This jet dominance also implies that 
\emph{all hard state sources should be radiatively inefficient}.
The reason for this inefficiency could be advection into the jet 
as well as advection into the black hole.
\section{Conclusions}
\label{sec:conc}
The puzzling optical/X-ray correlations of 
XTE~J1118+480, can be understood in terms of a common energy reservoir for
both the jet and the Comptonizing electrons.
Any energy reservoir model for XTE J1118+480
requires that the total jet power dominates
 over the X-ray luminosity. 
 Following the same line of arguments as FGJ03, we showed that
this situation is likely and probably represents a common feature of all
black holes in the low-hard state.

\end{article}
\end{document}